# Giant Thermal Magnetoresistance Driven by Graphene Magnetoplasmon

Ming-Jian He[1,2], Hong Qi[1,2,*], Yan-Xiong Su[1,2], Ya-Tao Ren[1,2], Yi-Jun Zhao[1], and Mauro Antezza[3,4]

1 School of Energy Science and Engineering, Harbin Institute of Technology, Harbin 150001, P. R. China

2 Key Laboratory of Aerospace Thermophysics, Ministry of Industry and Information Technology, Harbin 150001, P. R. China

3 Laboratoire Charles Coulomb (L2C), UMR 5221 CNRS-Université de Montpellier, F-34095 Montpellier, France

4 Institut Universitaire de France, 1 rue Descartes, F-75231 Paris, France

*Corresponding authors: Email: qihong@hit.edu.cn (H. Qi)

**Abstract:** In this work, we have predicted a giant thermal magnetoresistance for the thermal photon transport based on the tunable magnetoplasmon of graphene. By applying an external magnetic field, we find that the heat flux can be modulated by approximately three orders of magnitude. Accordingly, negative and giant relative thermal magnetoresistance ratios are both achieved for magnetic fields with a maximum strength of 4 Tesla. This effect is mainly caused by the suppression and enhancement of scattering interactions mediated by graphene magnetoplasmon. Specifically, it has never been achieved before for nanoparticles, which have no response to magnetic fields. The effect is remarkable at these reasonable strengths of fields, and thus has considerable significance for the real-life applications. It is also expected to enable technological advances for the thermal measurement-based magnetic sensor and magnetically thermal management.

The giant magnetoresistance effect discovered by Grünberg and Fert in 1988 [1] is considered as one of the most fascinating advances in solid state physics. Since then, extensive applications of magnetoresistance have been developed in electronics, such as magnetic sensors and hard-disk read-heads [2]. Inspired by the unique effect, a thermal analog named giant thermal magnetoresistance (GTM) effect is predicted in magneto-optical plasmonic structures in the context of radiative heat transfer [3]. Nowadays, the radiative heat transfer at the nanoscale is of great current interest [4-8] and has highlighted the possibility of modulating heat flux with entirely novel schemes [9-13]. Among them, the magneto-optical material InSb has showed remarkable performance in modulating the radiative heat transfer due to the magnetically tunable properties [5, 14-17]. More recently, by investigating the near-field radiative heat transfer between two InSb particles, a huge anisotropic thermal magnetoresistance with values of up to 800% is achieved with a magnetic field of 5 T [18]. The GTM effect has great application significance for the thermal measurement-based magnetic sensing and magnetically thermal management. However, the GTM in plasmonic structures has so far strictly relied on the magneto-optical nanoparticles made of semiconductors, like InSb. The realization of GTM is still demanding for nanoparticle structures made of conventional materials, which has no response to magnetic field.

By locating a substrate near two nanoparticles, Dong et al. [19] have introduced a new channel of propagating surface waves to assist the heat transfer. They have successfully delayed the deterioration of thermal photon transport, especially in long distance. Then the performance of the long-distance energy-exchange was improved by exciting the surface plasmon polaritons of graphene [8, 20]. Specifically, in the presence of an external magnetic field, hybridization occurs between cyclotron excitations and plasmons in graphene, originating magnetoplasmon polaritons (MPP) [21]. The MPP effect has been utilized to realize magnetically tunable near-field radiative heat transfer between suspended graphene sheets [22, 23] and gratings [24]. In the present work, based on the graphene MPP, we have successfully proposed a scheme to induce a GTM effect between two nanoparticles, made of common materials silicon dioxide ($SiO_2$). It should be mentioned that the giant thermal magnetoresistance effect obtained in the present work is indeed remarkable and much stronger than those in the previous study [3].

Here we consider two $SiO_2$ [25] nanoparticles located above a graphene sheet with a distance $z_n$, and they are separated with a distance $d$ as illustrated in Fig. 1. The radii of the two nanoparticles are identical and selected as 5 nm, which has been widely used in the previous studies [15, 20, 26]. We limit the calculations with the particle-surface distance $z_n$ = 50 nm and particle-particle distance $d \geq 100$ nm to guarantee the validity of the dipolar approximation [15, 20]. The two nanoparticles are kept at temperature $T_1$ and $T_2$. A static magnetic field with intensity $B$ is applied perpendicularly to the graphene sheet. It should be mentioned that, the substrate effect is ignored in the present work to avoid the MPP hybridization with other modes. Thus the pure effect of MPP on the heat transfer mechanism can be distinguished clearly. Moreover, suspended graphene sheets are always considered as ideal physical models for the studies of near-field radiative heat transfer [11, 22, 23, 27]. As illustrated in Fig. 1, the thermal photons transfer through two channels, (1) the direct particle-particle channel via vacuum interaction and (2) the particle-graphene-particle channel via scattering interaction.

With the application of an external magnetic field, a characteristic optical quantum Hall effect occurs to graphene electrons [28]. The conductivity of graphene becomes a tensor with nonzero elements in off-diagonal parts

$$\begin{pmatrix} \sigma_{xx} & \sigma_{xy} \\ \sigma_{yx} & \sigma_{yy} \end{pmatrix} = \begin{pmatrix} \sigma_L & \sigma_H \\ -\sigma_H & \sigma_L \end{pmatrix} \quad (1)$$

where $\sigma_L$ and $\sigma_H$ denote the longitudinal and Hall conductivities, respectively. The magneto-optical conductivities are taken from Refs. [28, 29] and the chemical potential of graphene is selected as $\mu = 0.08$ eV throughout the letter. The intensity of magnetic field is limited to $B = 0\sim4$ T, which has significance for the real-life applications. Under this circumstance, the effect of magnetic fields on modifying the optical properties of $SiO_2$ can be ignored [30]. Based on the dipole approximation, the electric polarizabilities of the nanoparticles are given as [20]

$$\alpha^{(0)}(\omega) = 4\pi R^3 \frac{\varepsilon(\omega)-1}{\varepsilon(\omega)+2} \quad (2)$$

where $R$ and $\varepsilon(\omega)$ are the radius and dielectric function of the nanoparticles, respectively. The polarizability needs to be modified by fluctuation-dissipation theorem

$$\chi(\omega) = \text{Im}[\alpha(\omega)] - \frac{k_0^3}{6\pi}|\alpha(\omega)|^2 \quad (3)$$

where $\alpha(\omega) = \alpha^{(0)}(\omega)/[1 - i\omega^3\alpha^{(0)}(\omega)/(6\pi c^3)]$ is the dressed polarizability with the radiation correction and $k_0 = \omega/c$.

Then we introduce the radiative heat transfer in the proposed system. The whole system is assumed to be thermalized at $T=T_1=T_2=300$ K at the initial state, then nanoparticle 1 is heated up to $T_1 = T + \Delta T$. This causes a heat flux $\varphi$ between the two nanoparticles. A radiative heat transfer conductance is defined to quantitatively evaluate the heat flux

$$h = \lim_{\Delta T \to 0} \frac{\varphi}{\Delta T} = 4\int_0^{+\infty} \frac{d\omega}{2\pi} \hbar\omega k_0^4 \frac{\partial n(\omega,T)}{\partial T} \chi^2 \mathrm{Tr}(\mathbf{GG}^*) \tag{4}$$

where $n(\omega,T) = \left[\exp(\frac{\hbar\omega}{k_B T}) - 1\right]^{-1}$ is the Bose-Einstein distribution and * denotes conjugate transpose. **G** is the dyadic Green tensor composed of two parts, i.e., $\mathbf{G} = \mathbf{G}^{(0)} + \mathbf{G}^{(sc)}$. $\mathbf{G}^{(0)}$ and $\mathbf{G}^{(sc)}$ represent the contributions of vacuum and scattering interaction, accounting for direct particle-particle and particle-interface-particle channels, respectively. The two parts read as

$$\mathbf{G}^{(0)} = \frac{e^{ik_0 d}}{4\pi k_0^2 d^3} \begin{pmatrix} a & 0 & 0 \\ 0 & b & 0 \\ 0 & 0 & b \end{pmatrix} \tag{5-a}$$

$$\mathbf{G}^{(sc)} = \int_0^{+\infty} \frac{dk}{2\pi} \frac{ike^{2ik_z z}}{2k_0^2 k_z} (r_s \mathbf{S} + r_p \mathbf{P}) \tag{5-b}$$

where $a = 2 - 2ik_0 d$, $b = k_0^2 d^2 + ik_0 d - 1$. $k$ and $k_z = \sqrt{k_0^2 - k^2}$ are the parallel and perpendicular wave-vectors. More details including the matrices **S** and **P** can be found in Refs. [20]. We note that due to the Hall conductivities, the cross-polarization reflection coefficients $r_{sp}$ and $r_{ps}$ are involved in the reflection characteristics of magneto-optical graphene [22-24]. They do not appear in Eq. (5-b), whereas it does not mean that the Hall conductivities play no role in the scattering interaction. In particular, the $r_s$ and $r_p$ in Eq. (5-b) is modified to take in consideration of the effects of both $\sigma_L$ and $\sigma_H$ [31]

$$r_s = \frac{2\sigma_L Z^h + \eta_0^2(\sigma_L^2 + \sigma_H^2)}{-(2 + Z^h \sigma_L)(2 + Z^e \sigma_L) - \eta_0^2 \sigma_H^2} \tag{6-a}$$

$$r_p = \frac{2\sigma_L Z^e + \eta_0^2(\sigma_L^2 + \sigma_H^2)}{(2 + Z^h \sigma_L)(2 + Z^e \sigma_L) + \eta_0^2 \sigma_H^2} \tag{6-b}$$

where $Z^h = i\omega\mu_0/k$, $Z^e = ik/\omega\varepsilon_0$, and $\eta_0$ denotes the free-space impedance.

A recent experimental study has shown that the radiative heat transfer between two coplanar membranes can be modulated by bringing a third planar object into close proximity [9]. The configuration is similar to the present work. However, with the determined geometric parameters in the system, the heat flux is fixed. This work presents a novel scheme for the configuration to dynamically modulate thermal transport with the magnetic method. In Fig. 2(a), for different separation distance $d$ between the two particles, the modulation performance of magnetic field is demonstrated by the modulation factors $\eta=h(B)/h(0)$ as the function of $B$. We show that by tuning the intensities of the magnetic fields, the heat transfer can be either enhanced or suppressed compared to that of zero field. The factor $\eta$ can be tuned nearly over three orders of magnitude with the reasonable strength of fields, $B$ = 0-4 T. In addition, we observe a slight oscillation of $\eta$ emerging at weak fields $B$ = 0-0.6 T. To reveal the GTM in the proposed system, a relative thermal magnetoresistance ratio is defined as $R_{TMR}=[R(B)-R(0)]/R(0)=[h(0)/h(B)-1]\times 100\%$, where the thermal magnetoresistance is given by $R=1/h$. As given by the definition, the positive and negative values of $R_{TMR}$ represent the decayed and enhanced heat transport compared to the zero field. It is shown in Fig. 2(a) that at $d$ = 1556 nm, $R_{TMR}$ reaches values of up to 635% and low to -83.7% at the field $B$ = 4 T and 1.17 T, respectively. For $d$ = 673 nm, the maximum and minimum of $R_{TMR}$ are achieved at $B$ = 4 T and 1.09 T for 7734% and -58.5%, respectively. It should be mentioned that the giant $R_{TMR}$ = 7734% is much larger than those of previous studies [3, 14, 18]. Additionally, the negative $R_{TMR}$ = -83.7% reveals relatively strong enhancement of heat transfer with reasonable strengths of the fields, which are much weaker than those of the existing studies on magnetically tunable radiative heat transfer [17, 22]. To make out the heat transfer mechanism accounting for the above GTM, we plot in Fig. 2(b) the scattering ratios defined as $\eta_S = h/h^{(0,0)}$, where $h^{(0,0)}$ denotes the contribution of vacuum interactions in the heat transfer conductance. A horizontal dashed line corresponding to $\eta_S$ = 1 is added in Fig. 2(b), which denotes the circumstance when scattering interactions vanish. The results demonstrate that scattering ratios $\eta_S$ in Fig. 2(b) and the modulation factors $\eta$ in Fig. 2(a) exhibit similar variety trends with $B$. As discussed above, the higher $\eta_S$ stands for more participation of scattering interaction, which is dominated by the graphene MPP. Despite of the slight oscillation at weak fields, the primary trend of $\eta_S$ is decaying with $B$. This phenomenon implies that a transition from scattering enhancement to scattering suppression occurs with the enlarging magnetic fields.

To explore the physical mechanism of the GTM, the radiative heat transfer conductance $h(B, \omega)$ is demonstrated in Fig. 3(a) for $d$=1556 nm. As investigated in previous studies, the phonon polaritons of $SiO_2$ are excited in the frequency ranges $8.67\times 10^{13}$~$9.47\times 10^{13}$ rad/s and $2.03\times 10^{14}$ ~$2.35\times 10^{14}$ rad/s [8, 32]. The two branches at the specific frequencies are also observed in Fig. 3(a), whereas they differ from each other sharply in magnetic-dependent spectrum. The low-frequency branch converts to broadband at weak fields due to the interactions with graphene MPP. Then, as $B$ enhances, this branch is divided into two parts. They correspond to the intraband (low-frequency) and the first interband transitions (high-frequency) of graphene MPP [24, 28], respectively. The maximum of $h(B, \omega)$ occurs at $B = 1.17$ T and $\omega = 0.92 \times 10^{14}$ rad/s, where the first interband MPP interacts with low-frequency phonon polaritons of $SiO_2$ and forms a strong coupling. To take a deep insight into the effect of the strengths of magnetic fields, we plot $h(\omega)$ in Fig. 3(b) corresponding to different cases: (1) the vacuum conductance without graphene, and (2) different strengths of fields, which are sliced with dashed lines in Fig. 3(a). The peak value of $h(\omega)$ for $B = 1.17$ T is larger than those of $B = 0, 2$ T and $B = 3, 4$ T by approximately one and two orders of magnitude. We can infer that this enhancement of $h(\omega)$ leads to the $R_{TMR} = -83.7\%$ in Fig. 2(a). In addition, we find that as $B$ enhances and exceeds 3 T, the intraband MPP almost fade out and the interband MPP decouple with phonon polaritons of $SiO_2$. This results in the decaying trend of $h(\omega)$ with $B$, and $h(\omega)$ nearly reduces to that of vacuum interaction with $B > 3$ T. The above results imply that the evolution of the MPP modes with magnetic fields plays a crucial role in GTM. Therefore in Fig. 3(c), we demonstrate the propagating properties of the MPP modes. They are given by the reciprocal of localization length in $x$ direction, defined as $l_x=q'/q''$ [21]. $q'$ and $q''$ represent the real and imaginary parts of complex longitudinal wave vector, respectively. A red dashed line is added in Fig. 3(c), indicating the same frequency with that in Fig. 3(b). The different peak values of $h(\omega)$ can be well explained by the magnitudes of $l_x$. We can confirm that the MPP modulates the scattering interactions based on the strong dependence of propagating length on the intensities of magnetic fields.

To have an intuitive understanding of the scattering characteristics, in Fig. 4(a), the electric field energy density $u_e$ [19] is illustrated in $x$-$y$ plane ($z = z_n/2$) for $\omega = 0.92 \times 10^{14}$ rad/s, corresponding to the peak in Fig. 3(b). $u_e$ of $B = 4$ T are considerably weakened compared to those of zero field and $B = 1.17$ T. In addition, compared to zero field, $u_e$ has enhanced when a magnetic field $B = 1.17$ T is applied. In Figs. 4(b) and 4(c), the ratios of $u_e$ in $B = 1.17$ T and $B = 4$ T to that in zero field are demonstrated in the $x$-$z$ plane. The two nanoparticles are indicated by red spots near $x = \pm 0.8$ $\mu$m. The positive and negative ratios represent the enhancement and suppression of $u_e$. A scattering enhancement occurs in Fig. 4(b) between the two nanoparticles, and it is concentrated above the graphene with a distance $z \approx 1.5$ $\mu$m. Interestingly, the strongest enhancement locates away from the graphene surface, and even higher than the nanoparticles. As is known, the surface plasmon polaritons of graphene are always excited and confined near the interface [21], and it is the same with the scattering enhancement by graphene [8]. The unique scattering enhancement at this position has never been observed in graphene plasmons before. Specifically, a remarkable scattering suppression is observed in Fig. 4(c) near the graphene sheet. The $B =$ 4 T field shuts off the energy-exchange channel dominated by graphene, and thus an attenuation works on the heat flux between the particles. Then, we can conclude from the above results that, the negative and giant thermal magnetoresistance effects are attributed to the MPP scattering enhancement and MPP scattering suppression, respectively.

In summary, we have predicted a negative and a giant thermal magnetoresistance effect between two $SiO_2$ nanoparticles based on the graphene MPP. The relative thermal magnetoresistance ratio can reach values of up to 7734% and low to -83.7% for a magnetic field of 4 T and 1.17 T, respectively. These values are indeed remarkable at these strengths of fields and have never been achieved in nanoparticle structures made of conventional materials, which has no response to magnetic field. We show that this behavior is mainly resulted from the suppression and enhancement of scattering interaction mediated by graphene MPP. The effect in the present work is promising for the thermal measurement-based magnetic sensing and magnetically thermal management. The physics in this work are limited to the fixed chemical potentials of graphene and geometry sizes ($R$, $z_n$). We believe that the optimized parameters can result in more considerable GTM effect. Moreover, we expect in the future work, magneto-optical materials like InSb can act as the substrate and assist the GTM.

## ACKNOWLEDGEMENTS

The supports of this work by the National Natural Science Foundation of China (No. 51976044, 51806047) are gratefully acknowledged. Heilongjiang Touyan Innovation Team Program is gratefully acknowledged. M. A. acknowledges support from the Institute Universitaire de France, Paris, France (UE).## DATA AVAILABILITY

The data that support the findings of this study are available from the corresponding author upon reasonable request.

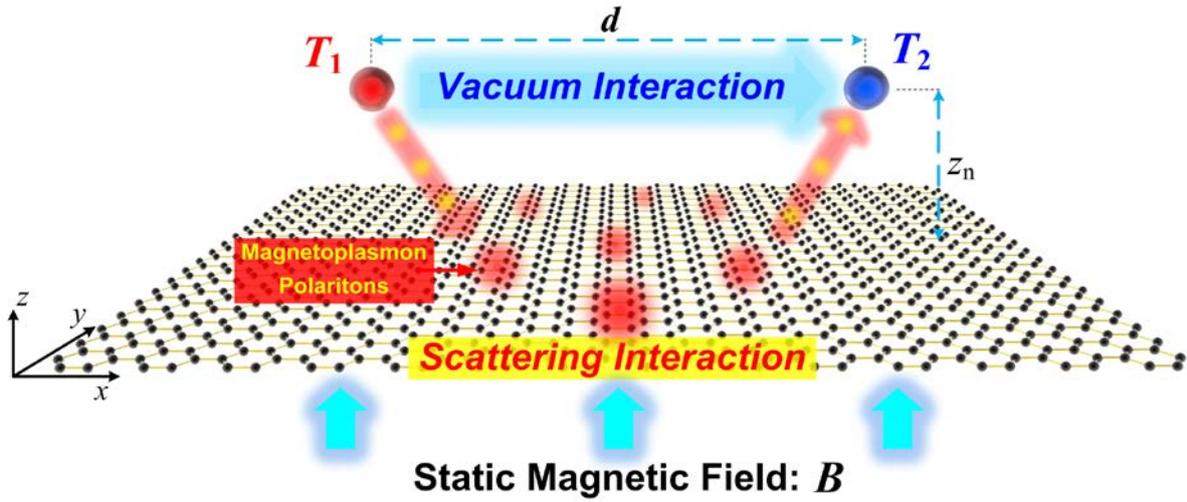

FIG. 1  Schematic of radiative heat transfer between two nanoparticles (separated with a distance *d*) located above a graphene sheet with a distance $z_n$. A static magnetic field with intensity *B* is applied perpendicularly to the graphene sheet. There are two channels for energy-exchange, (1) the direct particle-particle channel via vacuum interaction, and (2) the particle-graphene-particle channel via scattering interaction.

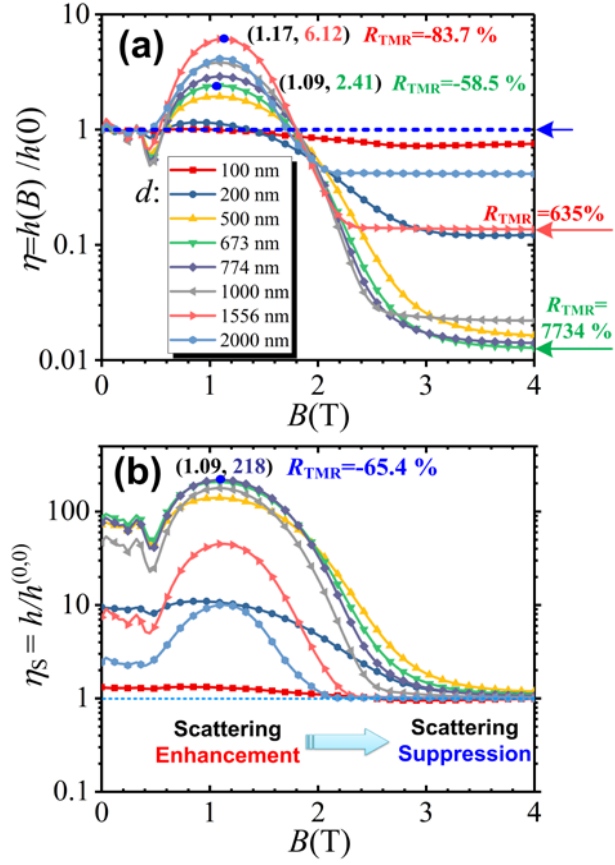

FIG. 2 (a) Modulation factors $\eta = h(B)/h(0)$ together with the relative thermal magnetoresistance ratio $R_{TMR} = [R(B)-R(0)]/R(0)=[h(0)/h(B)-1]\times 100\%$. (b) Scattering ratios defined as $\eta_S = h/h^{(0,0)}$, where $h^{(0,0)}$ denotes the contribution of vacuum interactions in the heat transfer conductance.

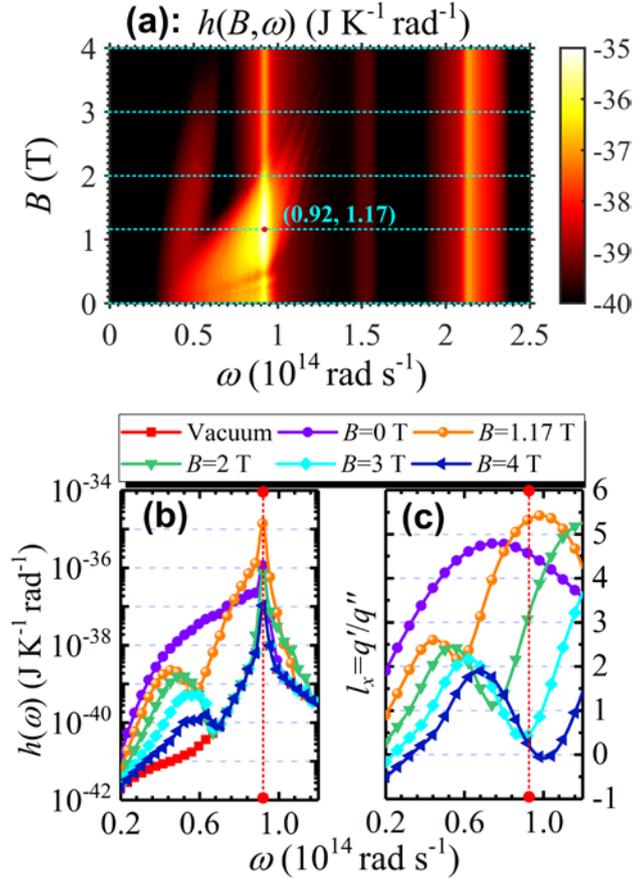

FIG. 3  (a) Radiative heat transfer conductance $h(B, \omega)$ for $d$=1556 nm. (b) Spectral heat transfer conductance $h(\omega)$ for vacuum interaction and different fields with scattering. (c) Propagating properties of the MPP modes as the reciprocal of localization length in $x$ direction.

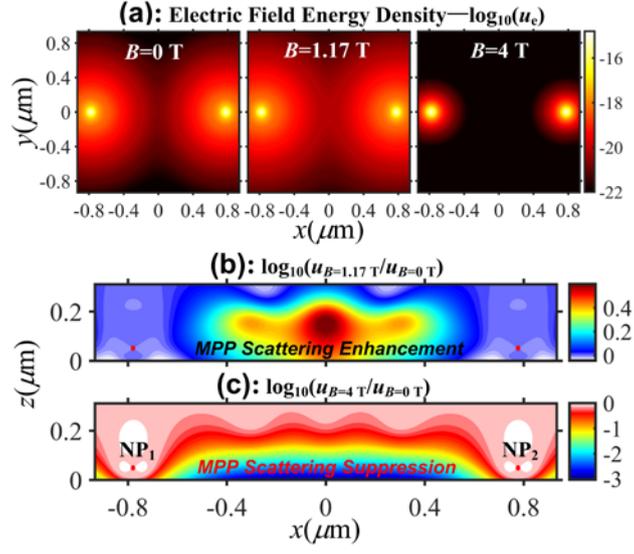

FIG. 4  (a) Electric field energy density $u_e$ in $x$-$y$ plane ($z = z_n/2$) for $\omega = 0.92\times10^{14}$ rad/s. Ratio of $u_e$ in (b) $B =$ 1.17 T and (c) $B = 4$ T to that in zero field in the $x$-$z$ plane. The two nanoparticles are indicated by red spots near $x = \pm 0.8$ μm.